\documentclass[aip,jcp,reprint,superscriptaddress,nofootinbib,floatfix]{revtex4-2}

\usepackage{amsmath,amssymb,bm,mathtools}
\usepackage{graphicx,booktabs,microtype,xcolor}
\usepackage[colorlinks=true,citecolor=blue!60!black,linkcolor=blue!60!black,urlcolor=blue!60!black]{hyperref}

\newcommand{\dd}{\mathrm d}
\newcommand{\E}{\mathbb E}
\newcommand{\Cov}{\operatorname{Cov}}
\newcommand{\Var}{\operatorname{Var}}

\begin{document}

\title{A Shared Observation Shields Collective Fluctuations while Preserving Local Independence}

\author{Hu Cang}
\email{cangh@uci.edu}
\affiliation{Department of Developmental and Cell Biology, University of California, Irvine, Irvine, California 92697, USA}

\date{\today}

\begin{abstract}
As a liquid approaches its glass transition, its dynamics turns heterogeneous: mobile and immobile regions coexist, and the four-point susceptibility $\chi_4$ that quantifies this heterogeneity grows sharply. Interpreting that growth is subtle, because the collective signals experiments record, such as a tagged particle's trajectory, an overlap function, or a mean field, are generated by the same particles they describe. Here we compute exactly what conditioning on such a shared record does to the population that produced it, for a broad class of stochastically observed systems; the guiding example is a tagged particle and the cage of $z$ neighbors that drives its force history. Using a Girsanov path transformation, we prove that the conditioning multiplies the independent joint law of the $z$ trajectories by exactly one term: a centered-square penalty along the single collective direction the record can see. Any fixed pair of particles stays nearly independent, with covariance falling as $O(z^{-1})$ and mutual information as $O(z^{-2})$, the property known as propagation of chaos, yet the $z(z-1)$ weak pair correlations add coherently into a finite suppression of collective fluctuations, the Schur shield $D - C = -C^2(aI + C)^{-1} \preceq 0$. An exactly solvable Brownian model calibrates the construction. The physical consequence is a calculable baseline for dynamical heterogeneity: conditioning itself contributes a computable, nonpositive amount to the susceptibility of a conditioned ensemble, so the genuine cooperative signal is the excess of the measured $\chi_4$ over this baseline rather than over zero, a comparison that existing simulation data can already perform.
\end{abstract}

\maketitle

\section{Introduction}\label{sec:intro}

Cool a liquid toward its glass transition and its average structure barely changes, but its dynamics breaks into bursts: compact groups of particles rearrange while nearby regions stay frozen \cite{Angell1995,Ediger2000,DebenedettiStillinger2001,BerthierBiroli2011}. Where this cooperative motion comes from is a central question of glass physics, because it appears without static order and without strong pair correlations.

Experiments and simulations measure the cooperativity with dynamic propensities, four-point susceptibilities $\chi_4(t)$, and correlated displacement fields \cite{BerthierEtAl2005,WidmerCooper2004,BerthierJack2007,TongTanaka2018}. Each of these observables sums fluctuations over a large population, and each hides the same subtlety: the collective coordinate is usually computed from the very trajectories it is meant to explain. Conditioning trajectories on a record they generated can produce a coherent population response even when the particles barely interact.

A tagged particle in a dense liquid is the guiding example throughout. The record $X$ is its measured trajectory, say a tracer colloid followed by confocal microscopy. The $z$ hidden paths are the neighbors that cage it, and the aggregate signal is the collective force the cage exerts on the tracer. These identifications fix language only; every statement below is a theorem about the model of Sec.~\ref{sec:model} and Appendix~\ref{app:model}.

Such a record is endogenous. The tracer feels drag from the very particles it reports on; an overlap function is computed from the paths it summarizes; a mean field is generated by the population that responds to it. Fixing the record forbids the collective excursions that would have changed it, while differences between one recorded history and the next align the particles that produced them, the way an audience falls into step by following one conductor rather than by watching each other.

What makes the interpretation subtle is a separation of scales. In a large exchangeable population, any two tagged members become statistically independent as the population size $z$ grows, a classical result known as propagation of chaos \cite{BenArousZeitouni1999,Lacker2021,Lacker2023}; watch two chosen particles and their correlation decays as $1/z$. A bulk susceptibility, however, sums over all $z(z-1)$ pairs, so correlations too small to detect pair by pair still add up to an order-one collective signal. Pair independence and collective heterogeneity therefore coexist. The question with practical consequences is which weak pair correlation survives the sum, and how much of the sum the observation itself supplies. Pair-independence bounds control fixed labels, and bulk fluctuation theories control full sums; neither separates these two mechanisms.

\subsection{Physical origin and historical development}

Two mathematical traditions frame the problem. One, running from Kac's and McKean's founding analyses of molecular chaos \cite{Kac1956,McKean1967} to modern quantitative bounds \cite{BenArousZeitouni1999,Lacker2021,Lacker2023}, proves that fixed pairs decouple, with covariance of order $1/z$. The other, developed by Ellis and Newman, Bolthausen, and Kusuoka and Liang, evaluates the fluctuations of population sums and shows that they stay finite \cite{EllisNewman1978,Bolthausen1986,KusuokaLiang2000}.

Recent mathematical work has begun to connect the two scales through conditional common noise, residual dependence, and entropy bounds \cite{Bally2026,Jalowy2024,BenArousZeitouni1999}. Still missing was the explicit signed interaction that appears when a population drives its own observation, and with it the answer to the question above: which weak pair correlation survives the collective sum.

This paper supplies that interaction. It also prices a standard theoretical step: mode-coupling and projection theories describe a tagged particle conditioned on the force history of its cage and then factorize the neighbors, and the cost of that conditioning is exactly what we compute. A Girsanov change of measure shows that conditioning on an endogenous record multiplies the independent path distribution by a single centered-square penalty, Eq.~\eqref{eq:posterior}. Everything else follows from this one term. The penalty acts only along the collective direction the record can see, and it suppresses rather than amplifies motion in that direction; we call the resulting operator inequality the Schur shield. Fixed pairs keep covariance of order $1/z$ and mutual information of order $1/z^{2}$, while the resummed susceptibility keeps an order-one negative correction. A sign threshold then separates the shield from the alignment created by history-to-history variability, and an exactly solvable model calibrates both.  The practical output is a baseline: interpreting a measured susceptibility requires computing the conditioning-only reference first, and reading heterogeneity as the excess over that reference, not over zero.

\subsection{Why the distinction matters}

Separating true pairwise interactions from observation-induced shielding matters wherever collective observables are built from trajectories. In glassy liquids, four-point susceptibilities and dynamic propensities quantify spatial heterogeneity \cite{WidmerCooper2004,BerthierEtAl2005,TongTanaka2018,BerthierJack2007,Bapst2020,Pezzicoli2024}, and time-domain optical spectroscopy follows the crossover from fast local motion to mode-coupling relaxation through one recorded collective signal \cite{Cang2003JCP,Cang2003CPL,Cang2005,Cang2003LC}. The theorem gives all of them one controlled baseline: the exact conditioned component that any analysis built on a shared record contains.

The calculation also speaks to two established bodies of theory. Projection formalisms and mode-coupling hierarchies separate tagged particles or collective order parameters from their environment \cite{Mori1965,Zwanzig1960,Gotze2009,Janssen2018}, with modern versions retaining multitime histories \cite{VanZon2002,Latz2000,Mayer2006,Janssen2015,Luo2020,Debets2021}; the theorem states exactly what conditioning on an aggregate output does to the remaining degrees of freedom, namely a rank-one penalty. Mathematical studies of pair decoupling \cite{Lacker2021,Lacker2023,Nikolaev2024,LiWangWang2025} and of bulk fluctuations \cite{EllisNewman1978,Bolthausen1986,KusuokaLiang2000,AlbeverioLiang2005} treat the two scales separately; keeping the explicit $1/z$ pair covariance is what connects them.

One conditional law therefore yields the pair covariance prefactor, its $1/z^{2}$ information cost, and the resummed susceptibility, and it cleanly splits fluctuation suppression at fixed history from variance generated across histories. Section~\ref{sec:model} derives the penalty and the shield. Section~\ref{sec:chaos} computes the pair covariance and resums it. Section~\ref{sec:hetero} lets the record fluctuate and derives the sign threshold. Section~\ref{sec:gaussian} evaluates everything exactly in a solvable dynamical model, and Sec.~\ref{sec:implications} states what to measure.

\begin{figure*}[t]
\centering
\includegraphics[width=0.7\textwidth]{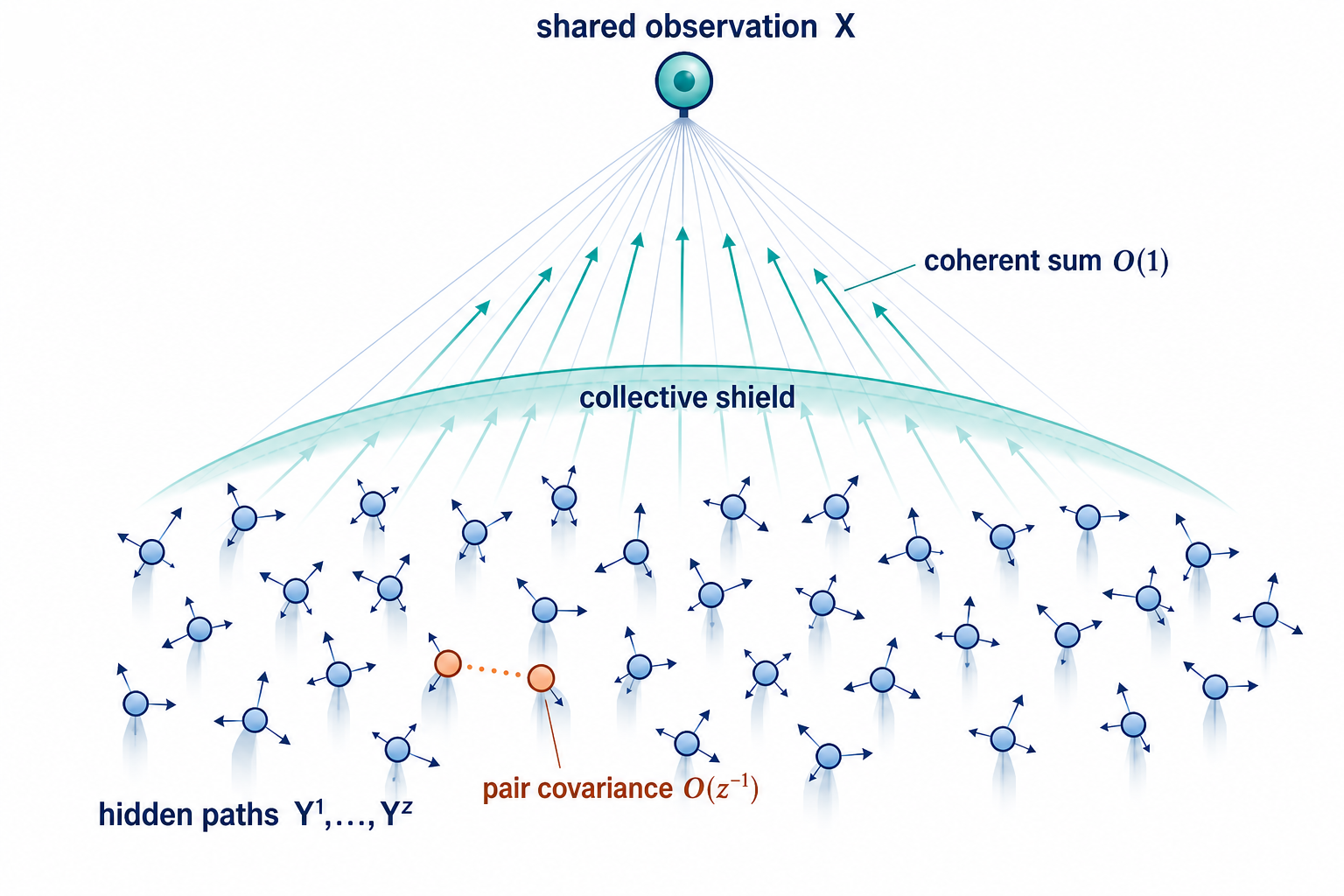}
\caption{One shared output constrains a whole population.  The diagram is schematic and has no coordinate axes.  In the tagged-particle reading, $X$ is the measured tracer trajectory and the hidden paths $Y^1,\ldots,Y^z$ are its caging neighbors.  At fixed history the paths retain only $O(z^{-1})$ pair covariance, but their coherent sum remains $O(1)$.  Conditioning on the endogenous observation $X$ compresses that source-visible collective direction, represented by the translucent shield.}
\label{fig:picture}
\end{figure*}

\section{A finite-population path model}\label{sec:model}

Strip the problem to its mechanism. Take $z$ hidden trajectories whose average drives one noisy measured path. Fix that path, and collective deviations that would have altered it become unlikely. The dependence this creates should live in the population direction the record can see, not in an arbitrary network of pair interactions. The change of measure below confirms this.

Let $Y^1,\ldots,Y^z$ be hidden stochastic paths and let $X$ be a noisy aggregate output driven by their collective contribution.  In the Brownian realization that anchors all statements below (Appendix~\ref{app:model}), the reference law evolves the record on its own, $\dd X_t=b_0(X_t)\,\dd t+\sigma_0\,\dd B_t$, with the hidden paths independent given $X=x$, and the endogenous law adds the collective drift
\begin{equation}
 \dd X_t=b_0(X_t)\,\dd t
 +\frac{1}{\sqrt z}\sum_{i=1}^z f(X_t,Y_t^i)\,\dd t
 +\sigma_0\,\dd B_t,
 \label{eq:sde}
\end{equation}
with $f$ bounded, all coefficients Lipschitz, and a fixed observation window $[0,T]$.  The $z^{-1/2}$ normalization is the fluctuation scale of a sum of $z$ weakly coordinated contributions; it keeps the collective force comparable to the output noise as $z$ grows.  A Girsanov transformation \cite{Girsanov1960} from the reference dynamics to the endogenous dynamics gives the conditional posterior in the closed form
\begin{multline}
 \frac{\dd P_z^x}{\dd Q_x^{\otimes z}}(y_1,\ldots,y_z)
 =\frac{1}{Z_z(x)}\\
 {}\times\exp\!\left[-\frac{1}{2a z}
 \left\lVert\sum_{i=1}^z\bigl(G_x(y_i)-m_x\bigr)\right\rVert^2\right].
 \label{eq:posterior}
\end{multline}
Here $Q_x$ is a one-path tilted law, $G_x(y)(t)=f(x_t,y_t)$ is the path contribution to the observed drift, $m_x=\E_{Q_x}G_x$, $a=\sigma_0^2$ is the output-noise variance, and the norm is taken in the observation Hilbert space $L^2([0,T])$.  Every one-body effect is absorbed into $Q_x$.  The entire many-body interaction is the single centered square.  All remainders quoted below are nonasymptotic and dimension free at this fixed window, with constants depending only on the force bound, the window, and $a$; they are not uniform on a window that grows with the relaxation time (Sec.~\ref{sec:gaussian}).

Each object in Eq.~\eqref{eq:posterior} has a direct physical reading.  The law $Q_x$ describes one hidden path responding to the recorded history.  The function $G_x$ converts that path into its predicted contribution to the record.  The centered sum measures how far the population would push the output away from the value already conditioned on, and $a$ sets the tolerance: a noisy output forgives collective deviations, a precise output penalizes them.  The minus sign expresses a constraint, not a force.  The population cannot collectively move a signal that has already been fixed.

The interaction is rank structured, far simpler than an all-to-all coupling.  Microscopic motion that leaves $G_x$ unchanged still follows the independent one-path law to leading order; only the source-visible direction is screened.  This predicts anisotropic suppression: two observables with the same single-particle variance can pick up very different collective corrections depending on how strongly they overlap with the measured output.  The representation holds on a separable Hilbert space, so time-dependent observation needs no discretization of the path.

Define the centered observation variable $X_i=G_x(Y^i)-m_x$ under $Q_x$, with covariance operator
\begin{equation}
 C=\E_{Q_x}[X_i\otimes X_i].
\end{equation}
For the normalized sum $S_z=z^{-1/2}\sum_iX_i$, the conditional covariance has the large-$z$ limit
\begin{equation}
 \Cov_{P_z^x}(S_z)=D+O(z^{-1}),
 \qquad D=aC(aI+C)^{-1}.
 \label{eq:D}
\end{equation}
Since
\begin{equation}
 D-C=-C^2(aI+C)^{-1}\preceq0,
 \label{eq:shieldop}
\end{equation}
the observation reduces the visible collective fluctuations in every direction.  We call this operator difference the Schur shield.

Diagonalizing $C$ makes the shield concrete.  An eigenmode of variance $c$ ends up with conditioned variance $ac/(a+c)$: weakly observed modes with $c\ll a$ barely change, while strongly visible modes with $c\gg a$ saturate at the observation-noise scale.  This is the rule that combines two resistors in parallel, applied mode by mode.  It also shows that the shield can never overshoot: it subtracts a positive component while the total response stays positive.

\begin{figure*}[t]
\centering
\includegraphics[width=0.94\textwidth]{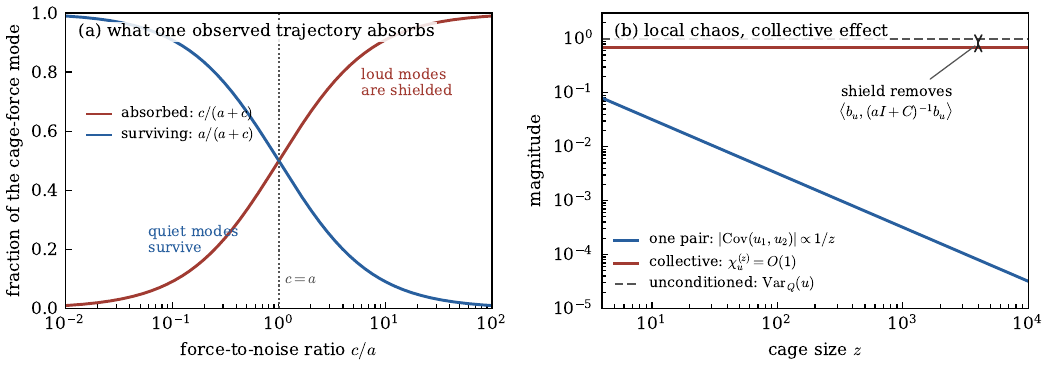}
\caption{A shared output selectively absorbs collective motion.  (a) An eigenmode of one-path variance $c$ splits into an absorbed share $c/(a+c)$ and a surviving share $a/(a+c)$ as its force-to-noise ratio $c/a$ grows.  (b) At fixed history, a tagged pair decorrelates as $1/z$ with the cage size $z$, the number of neighbors driving the record, while the normalized population sum retains the order-one shield $\langle b_u,(aI+C)^{-1}b_u\rangle$.  The dashed line shows the unconditioned collective scale.}
\label{fig:shield}
\end{figure*}

\section{Pair independence with a collective correction}\label{sec:chaos}

A local observer and a collective observer compute different sums.  The local observer fixes two labels while $z$ grows and should find independence.  The collective observer adds the same weak cross-correlation $z(z-1)$ times and can keep a finite response.  One calculation exposes both limits: remove two tagged paths from the collective constraint, then put their leading covariance back into the population sum.

Removing two labels and integrating over the other $z-2$ paths leaves a two-body log density of order $1/z$ with a centered leading term.  For bounded one-site observables $u$ and $v$,
\begin{multline}
 \Cov_{P_z^x}\bigl(u(Y^1),v(Y^2)\bigr)
 =-\frac{1}{z}\\
 {}\times\left\langle b_u,(aI+C)^{-1}b_v\right\rangle
 +O(z^{-3/2}),
 \label{eq:paircov}
\end{multline}
where
\begin{equation}
 b_u=\Cov_{Q_x}\bigl(u(Y),G_x(Y)\bigr).
\end{equation}
The pair mutual information is smaller, of order $1/z^{2}$, because relative entropy starts at the quadratic term in the small centered density perturbation.

Covariance and information therefore decay at different rates for the same pair.  A signed observable responds linearly to the $1/z$ density perturbation; mutual information discards the sign and starts quadratically.  A small pair-information estimate is thus no bound on a coherently summed susceptibility.  The object worth measuring is the covariance projected on the output-coupled direction, not a scalar dependence score.

Summing restores the collective view.  For the normalized population observable $z^{-1/2}\sum_i u(Y^i)$, Eq.~\eqref{eq:paircov} resums to
\begin{equation}
 \chi_u^{(z)}=\Var_{Q_x}(u)
 -\left\langle b_u,(aI+C)^{-1}b_u\right\rangle
 +O(z^{-1/2}).
 \label{eq:suscept}
\end{equation}
The second term is nonpositive, and it contains no adjustable parameter once the one-path law is known.  Equations~\eqref{eq:paircov} and \eqref{eq:suscept} hold simultaneously: the local observer and the collective observer probe different sums of the same weak correlation field.

The bookkeeping is elementary.  In the variance of $z^{-1/2}\sum_i u(Y^i)$, the $z$ diagonal terms give the one-path variance after normalization, and the $z(z-1)$ off-diagonal terms each carry a factor $1/z$, so the normalization leaves an order-one correction.  Pair independence controls any fixed set of labels, but it never justified exchanging the large-$z$ limit with this coherent double sum.

\section{Fluctuating histories and dynamical heterogeneity}\label{sec:hetero}

A fixed history is only one layer of a dynamical experiment.  Repeat the experiment and the observed history $X$ itself varies, and each history shifts the conditional mean, or propensity, of a tagged observable: $m_z(X)=\E[u(Y^1)\mid X]$.  Propensity fluctuations enter on the collective scale through $h_z(X)=\sqrt z\,[m_z(X)-\E m_z(X)]$; for the normalized collective observable $\widetilde U_z=z^{-1/2}\sum_i[u(Y^i)-\E u]$, exchangeability gives $\E[\widetilde U_z\mid X]=h_z(X)$.  Let $v_X$ be the single-path conditional variance and let
\begin{equation}
 \delta_X=\left\langle b_X,(aI+C_X)^{-1}b_X\right\rangle\geq0
\end{equation}
be the Schur shield for that history.  The law of total variance gives
\begin{equation}
 \Var(\widetilde U_z)
 =\E[v_X-\delta_X]+\Var(h_z(X))+O(z^{-1/2}).
 \label{eq:totalvar}
\end{equation}
For two different labels,
\begin{equation}
 z\,\Cov(u_1,u_2)
 =\Var(h_z(X))-\E\delta_X+O(z^{-1/2}).
 \label{eq:threshold}
\end{equation}

Equation~\eqref{eq:threshold} sets up a competition.  A shared fixed observation suppresses redundant collective motion, whereas variation between observed histories aligns particles through a common propensity.  The measured cross-correlation changes sign when the propensity variance overtakes the mean shield.  This is a general conditioned-population susceptibility, not an identification with every conventional bulk $\chi_4$ protocol; it supplies one controlled component that can be matched to a chosen experimental ensemble.

The decomposition explains a familiar frustration.  Two experiments can report opposite cross-particle signs while agreeing on the microscopic dynamics, simply because they condition on different information.  A protocol that fixes the macroscopic history weights the negative shield; a protocol that mixes histories adds the positive propensity variance.  Reporting the two terms of Eq.~\eqref{eq:threshold} separately, rather than only their sum, translates data between protocols.

\begin{figure}[t]
\centering
\includegraphics[width=\columnwidth]{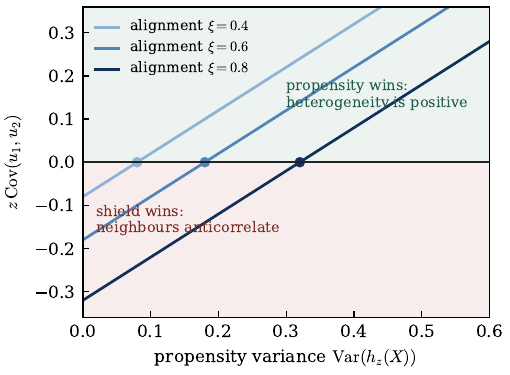}
\caption{Two sources set the sign of the cross-particle response.  The shared-output shield produces anticorrelation below each line; history-to-history propensity variance produces positive correlation above it.  Curves are labeled by the alignment $\xi\in[0,1]$, the normalized overlap between the observable's force coupling $b_u$ and the observed collective mode, for which the scalar shield is $\E\delta_X=\xi^2\Var(u)\,c/(a+c)$.  The crossing implements Eq.~\eqref{eq:threshold} without an adjustable sign convention.}
\label{fig:threshold}
\end{figure}

The propensity here is the path-space analogue of a familiar simulation tool.  For a two-time cage overlap, $m_z(X)$ is a conditional mobility propensity, with the observed history playing the role that the fixed initial configuration plays in the isoconfigurational ensemble \cite{WidmerCooper2004,TongTanaka2018}.  Equation~\eqref{eq:threshold} then makes a quantitative prediction: history-to-history propensity fluctuations must exceed the force-aligned shield before cross-particle dynamic heterogeneity turns positive.  Repeated runs from stored configurations, or any other conditional ensemble, can test this threshold without new experiments.

\begin{figure}[t]
\centering
\includegraphics[width=\columnwidth]{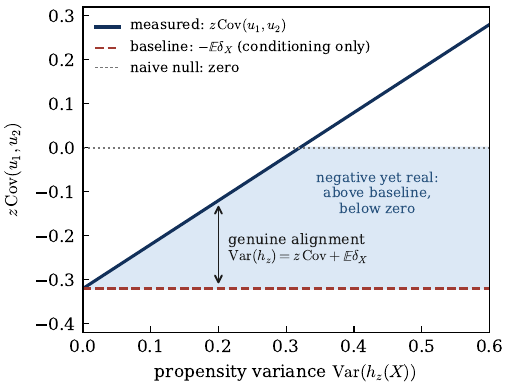}
\caption{Reading a susceptibility against the correct baseline.  The conditioning-only reference for the cross-particle response is the negative baseline $-\E\delta_X$ (dashed), not zero (dotted).  Genuine alignment is the excess of the measurement over the baseline, $\Var(h_z)=z\,\Cov(u_1,u_2)+\E\delta_X$ (arrow).  In the shaded band a measurement is negative yet sits above the baseline, so it still contains positive genuine alignment that a comparison with zero would miss.  Parameters as in Figs.~\ref{fig:shield} and \ref{fig:threshold}: $\Var(u)=1$, $c/a=1$, $\xi=0.8$, so $\E\delta_X=0.32$.}
\label{fig:baseline}
\end{figure}

Equation~\eqref{eq:threshold} also dictates how to read a measurement (Fig.~\ref{fig:baseline}).  The conditioning-only reference for the cross-particle susceptibility is not zero but the negative baseline $-\E\delta_X$.  A measured $z\,\Cov(u_1,u_2)$ that is small or even negative can therefore still contain genuine alignment: the propensity variance is the excess of the measurement over the baseline, $\Var(h_z(X))=z\,\Cov(u_1,u_2)+\E\delta_X+O(z^{-1/2})$.  Comparing with zero instead of the baseline underestimates the genuine component by exactly the shield.

Ensemble and response analyses of dynamic fluctuations ask how a bulk liquid builds a growing correlated volume \cite{BiroliBouchaud2004,Biroli2006,Berthier2007,Franz2012}.  Equation~\eqref{eq:threshold} identifies one contribution that the observation protocol itself adds or removes, and it tells the experimentalist exactly which two numbers to report: the fixed-history shield and the between-history propensity variance.

\section{Dynamic Gaussian calibration}\label{sec:gaussian}

The operator formulas distinguish visibility at an instant from visibility accumulated over a history.  A solvable Gaussian process checks that this distinction is dynamical rather than notational: the same observation constraint passes through two relaxation filters, and the equal-time and time-integrated responses come out different.

The model makes every symbol concrete.  On the whole time line, let the hidden coordinates be independent stationary Ornstein--Uhlenbeck processes and let the record integrate their collective pull,
\begin{align}
 \dd y_i&=-\lambda y_i\,\dd t+\sqrt{2\lambda\sigma^2}\,\dd W_i,\nonumber\\
 \dd x&=\Bigl[-a_0x+\frac{g}{\sqrt z}\sum_{i=1}^z y_i\Bigr]\dd t
 +\sqrt{2D_0}\,\dd W_0,
 \label{eq:OUsde}
\end{align}
with independent Brownian motions, rates $a_0,\lambda>0$, coupling $g$, hidden stationary variance $\sigma^2$, and output-noise intensity $D_0$.  Conditioning on the complete stationary $x$ history is a classical Gaussian smoothing problem, so the posterior is Gaussian and every operator can be evaluated exactly (Appendix~\ref{app:frequency}).  The cross covariance of two hidden paths takes the form
\begin{align}
 \Cov[y_i(t),y_j(s)\mid x]
 &=-\frac{\sigma^2}{z}
 \left[e^{-\lambda|t-s|}-\frac{\lambda}{\alpha}
 e^{-\alpha|t-s|}\right],\nonumber\\
 \alpha^2&=\lambda^2+\frac{g^2\lambda\sigma^2}{D_0}.
 \label{eq:OU}
\end{align}
The formula confirms the $1/z$ pair scale and the negative sign at fixed history, and it separates the two responses: the predictable fraction is $1-\lambda/\alpha$ at equal time but $1-\lambda^2/\alpha^2$ after time integration.  No single static coupling reproduces both numbers.  Temporal filtering is part of the physical susceptibility.

The two rates have plain meanings.  The rate $\lambda$ is the intrinsic memory loss of one hidden coordinate, while $\alpha$ folds in the precision and strength of the aggregate observation.  Their difference sets how long two paths stay anticorrelated after the record is fixed.  Equal-time measurements sample the instantaneous overlap of the two filters, integrated response weights their lifetimes, and the comparison distinguishes a genuinely dynamical shield from a static covariance constraint.

The theorem is uniform on every fixed observation window.  Carrying it to windows that grow with a structural relaxation time requires an additional mixing or dissipativity estimate.  This boundary is itself informative: it names the input that a slow-relaxation extension must measure.

\section{Implications}\label{sec:implications}

Conditioning on a shared record disturbs nothing: the dynamics runs exactly as before, and no force is applied.  What changes is the ensemble.  Among all trajectories the population could follow, only those consistent with the record remain, and Eq.~\eqref{eq:posterior} states exactly how this selection reweighs them---a classical, purely statistical counterpart of measurement back-action.  Raising the output precision suppresses fluctuations along the visible collective mode through $(aI+C)^{-1}$ and leaves orthogonal modes alone.  Two experimental knobs emerge, each tunable on its own: the precision of the shared output and the diversity of sampled histories.

The result also corrects an inference that local measurements invite.  Pair information of order $1/z^{2}$ does not bound the collective response, because $z$ weak signed correlations aligned along one direction survive the sum.  Wherever an aggregate record is driven by the constituents that produce it, including probe trajectories in crowded media, coarse-grained order parameters, neural population recordings, and latent-variable models \cite{NoeNueske2013}, varying the measurement noise gives a direct experimental handle: output-visible susceptibilities scale with the shield factor, orthogonal ones do not.

For glass physics the payoff is a baseline with a definite sign.  The fixed-history susceptibility of Eq.~\eqref{eq:suscept} is the conditioned analogue of the four-point susceptibility $\chi_4(t)$, and conditioning can only subtract from it.  Interpreting dynamical heterogeneity therefore takes two steps: compute the conditioning-only baseline from the one-path law, then read the measured susceptibility as an excess over that baseline (Fig.~\ref{fig:baseline}).  Growth of dynamic heterogeneity on approach to arrest cannot originate in the conditioned sector.  It must exceed the negative shield through genuine dynamical correlations, such as recollisions, conserved fields, or critical amplification, and Eq.~\eqref{eq:threshold} turns this statement into a quantitative threshold against propensity fluctuations that existing simulation data can already test.  The theorem covers one tagged population exactly; carrying it to the bulk $\chi_4$, which compares fluctuations across different tagged stars, requires controlling cross-population correlations and stands as a concrete, well-posed open problem.

\begin{acknowledgments}
This work was supported by the National Human Genome Research Institute of the National Institutes of Health under Award No.~1R01HG014004-01.
\end{acknowledgments}

\section*{Author Declarations}

\subsection*{Conflict of Interest}
The author has no conflicts to disclose.

\section*{Data Availability}
Data sharing is not applicable to this article, as all results are analytical; the derivations supporting the findings are contained within the article and its appendices.

\appendix

\section{The stochastic model and the centered square}\label{app:model}

Let $B,W_1,\ldots,W_z$ be independent Brownian motions.  Under the reference law, the system evolves as
\begin{align}
 \dd X_t&=b_0(X_t)\,\dd t+\sigma_0\,\dd B_t,\nonumber\\
 \dd Y_t^i&=c(Y_t^i,X_t)\,\dd t+\sigma\,\dd W_t^i,
 \qquad 1\leq i\leq z,
\end{align}
with $b_0$ and $c$ globally Lipschitz, $\sigma_0,\sigma>0$, $|f|\leq F_\ast$, hidden initial data independent and identically distributed given $X$, and a fixed window $[0,T]$; the observation Hilbert space is $L^2([0,T])$.  The endogenous law changes only the observed equation, adding the collective drift of Eq.~\eqref{eq:sde}.  Girsanov's theorem \cite{Girsanov1960} gives the density
\begin{multline}
 \exp\biggl\{\frac{1}{\sigma_0^2}\int_0^T
 \frac{1}{\sqrt z}\sum_{i=1}^zf(X_t,Y_t^i)\,
 [\dd X_t-b_0(X_t)\dd t]\\
 -\frac{1}{2\sigma_0^2}\int_0^T
 \Bigl[\frac{1}{\sqrt z}\sum_{i=1}^zf(X_t,Y_t^i)\Bigr]^2\dd t\biggr\},
\end{multline}
and boundedness of $f$ gives Novikov's condition, so the density is well defined.  The linear term is a sum of one-label factors.  The quadratic term equals
\begin{equation}
 -\frac1{2\sigma_0^2z}\left\|
 \sum_{i=1}^zG_X(Y^i)\right\|^2,
\end{equation}
and the $z^{-1/2}$ drift normalization is what makes this coefficient $1/(2az)$ with $a=\sigma_0^2$; an empirical-average drift $z^{-1}\sum_if$ with order-one output noise would give a $z^{-2}$ coefficient instead, and no order-one shield would survive.  Completing the square around the one-path mean $m_X$ moves every remaining one-body term into the tilted law $Q_X$ and all constants into the normalization.  The only term that still couples different labels is
\begin{equation}
 -\frac1{2az}\left\|
 \sum_{i=1}^z\bigl(G_X(Y^i)-m_X\bigr)\right\|^2.
\end{equation}
Normalization gives Eq.~\eqref{eq:posterior}.  The derivation explains the sign: once the shared output is fixed, simultaneous motion of all hidden paths along the visible direction receives a quadratic cost.  Constants in every remainder quoted in the main text depend only on $(F_\ast,T,a)$ and are dimension free.

\section{Leave-two-out expansion}

Write $X_i=G_X(Y^i)-m_X$ and remove labels 1 and 2 from their sum:
\begin{equation}
 W_{z-2}=\frac1{\sqrt z}\sum_{i=3}^zX_i.
\end{equation}
Conditioned on $Y^1,Y^2$, the remaining integral is the Laplace transform of $W_{z-2}$ evaluated at a field of size $z^{-1/2}(X_1+X_2)$.  A second-order expansion around zero gives a two-label density of the form
\begin{equation}
 1-\frac1z\left\langle X_1,(aI+C)^{-1}X_2\right\rangle
 +O_{L^1}(z^{-3/2}).
\label{app:two}
\end{equation}
The one-body pieces are absorbed into the corrected marginals.  Multiplying Eq.~\eqref{app:two} by centered observables $u(Y^1)$ and $v(Y^2)$ proves Eq.~\eqref{eq:paircov}.

The same normal form explains the entropy rate.  The leading density perturbation is centered and has size $z^{-1}$.  Relative entropy starts at its quadratic term, so
\begin{equation}
 I(Y^1;Y^2\mid X)=O(z^{-2}).
\end{equation}
Covariance is linear in the perturbation and remains $O(z^{-1})$.

\section{Collective resummation}

Let $U_z=z^{-1/2}\sum_i(u(Y^i)-\E u)$.  Exchangeability gives
\begin{equation}
 \Var(U_z)=\Var(u)+(z-1)\Cov(u_1,u_2).
\end{equation}
Substitute Eq.~\eqref{eq:paircov}.  The factor $z-1$ cancels the pair scale $z^{-1}$ and yields Eq.~\eqref{eq:suscept}.  The correction is finite because all weak pair covariances point along the same observed direction.

For the observation variable itself, Gaussian resummation gives
\begin{equation}
 D^{-1}=C^{-1}+a^{-1}I
\end{equation}
on the range of $C$.  This is the familiar precision-addition rule: independent prior fluctuations and observation noise add at the inverse-covariance level.  Rearrangement gives Eq.~\eqref{eq:D}.

\section{Random histories}

Let $\mathcal X$ denote the sigma algebra generated by the observed history.  For the normalized collective observable $\widetilde U_z=z^{-1/2}\sum_i[u(Y^i)-\E u]$, the law of total variance states
\begin{equation}
 \Var(\widetilde U_z)=
 \E[\Var(\widetilde U_z\mid\mathcal X)]
 +\Var(\E[\widetilde U_z\mid\mathcal X]).
\end{equation}
The first term contains the one-site variance minus the Schur shield.  In the second, exchangeability gives $\E[\widetilde U_z\mid\mathcal X]=\sqrt z\,[m_z(X)-\E m_z(X)]=h_z(X)$: the $\sqrt z$ amplification of the propensity fluctuation is forced by the normalization of $\widetilde U_z$, not chosen.  This proves Eq.~\eqref{eq:totalvar}.  Applying the same decomposition to two distinct labels gives Eq.~\eqref{eq:threshold}.

This identity also marks the scope of comparison with a bulk four-point susceptibility.  Different ensembles hold different global variables fixed, and those constraints move terms between the two parts of total variance.  A comparison should therefore match the observed history, normalization, and fixed variables before comparing amplitudes.

\section{Gaussian calibration in frequency space}\label{app:frequency}

For the linear model of Eq.~\eqref{eq:OUsde}, Fourier transformation diagonalizes the path covariance.  Observing the complete $x$ history is equivalent to observing $(\partial_t+a_0)x=g\,S+\sqrt{2D_0}\,\xi_0$ with $S=z^{-1/2}\sum_iy_i$ and white noise $\xi_0$, so the record's own relaxation rate $a_0$ drops out of every conditional covariance.  Each temporal frequency $\omega$ carries the one-site spectrum
\begin{equation}
 C(\omega)=\frac{2\lambda\sigma^2}{\omega^2+\lambda^2},
\end{equation}
whose integral over $\omega/2\pi$ recovers the stationary variance $\sigma^2$.  Conditioning adds the observation precision $g^2/2D_0$ at every frequency, so the shielded spectrum is
\begin{equation}
 D(\omega)=\frac{aC(\omega)}{a+C(\omega)},
 \qquad a=\frac{2D_0}{g^2},
\end{equation}
with poles at $\pm i\alpha$, $\alpha^2=\lambda^2+g^2\lambda\sigma^2/D_0$.  Inverse Fourier transformation produces the two exponential scales in Eq.~\eqref{eq:OU}.  Evaluating at zero lag gives the equal-time predictable fraction $1-\lambda/\alpha$.  Evaluating at zero frequency gives the integrated fraction $1-\lambda^2/\alpha^2$.

The difference between these numbers is physically useful.  It shows that an observation can strongly constrain slow collective motion while leaving faster fluctuations comparatively free.  Any reduced model that replaces the path operator by one static number should state which temporal filter that number calibrates.

\section{Suggested measurement workflow}

A measurement can isolate the two contributions in Fig.~\ref{fig:threshold} and construct the baseline comparison of Fig.~\ref{fig:baseline}.
\begin{enumerate}
\item Repeat the experiment over many observed histories and divide each history into independent hidden-particle or replica samples.
\item Within each history, estimate the pair covariance and multiply it by $z$.  This gives the fixed-history shield up to the stated finite-$z$ remainder.
\item Across histories, estimate the variance of the conditional mean.  This is the propensity contribution.
\item Report their difference together with the total collective variance.  The identity in Eq.~\eqref{eq:totalvar} provides an internal consistency check.
\item Repeat with two temporal filters.  A change between equal-time and integrated results measures the path-space character seen in Eq.~\eqref{eq:OU}.
\end{enumerate}
This workflow turns the sign threshold into two separately observable quantities and avoids assigning a small net signal to a single mechanism.

\bibliographystyle{aipnum4-2}
\bibliography{references,cang_references}

\end{document}